\documentclass{icrc}

\usepackage{times}
 \usepackage{graphicx} 
\usepackage{eucal}

\begin{document}

\title{Bayesian event reconstruction and background rejection in
neutrino detectors}

\author[1]{Gary C. Hill}
\affil[1]{Department of Physics\\
           University of Wisconsin, Madison}

\correspondence{Gary C. Hill\\ ghill@alizarin.physics.wisc.edu}
\firstpage{1}
\pubyear{2001}
\maketitle

\begin{abstract}
Several large volume, high energy neutrino detectors are in operation or
in the design stage. Upward going signal neutrino events must be separated
from large backgrounds of downgoing cosmic ray induced atmospheric
 muons. To this end, a
Bayesian extension of the traditional maximum likelihood reconstruction
method will be described. Further, it will be shown how signals can be
separated from backgrounds through integration of Bayesian posterior
probability densities.
\end{abstract}
\section{Introduction}
Recent years have seen the planning, construction and
 operation of the first large volume, high energy
neutrino detectors. The pioneering work of the DUMAND collaboration was 
followed by the successful deployment and operation of the Lake Baikal and
AMANDA experiments, and the planning and design studies for the ANTARES and
Icecube detectors.  
These detectors use large arrays of
 optical modules to detect the Cerenkov radiation
from upward moving muons that result from neutrino interactions that take place 
in the surrounding water/ice or rock. These signals must be 
separated from a very large
 background of downward moving atmospheric muons. At the trigger level,
 the ratio of downgoing muons to expected upgoing neutrino-induced
 muons can range from $\sim 10^9$ for a surface detector,
 down to $\sim 10^5$ for a deep underwater/ice detector. 

\begin{sloppypar}
What we will call the ``conventional'' method of separating background 
   and signal is to use a maximum likelihood method to fit a muon track
   arrival direction (based on the observed photon arrival times and 
densities in the optical module array) followed by the application of quality cuts to reduce
   the number of mis-reconstructed background events (fakes), while   
   keeping a large fraction of signal.
Once these data are reduced to 
a manageable size, they are compared to Monte Carlo expectations for both
signals and backgrounds, and the origins of the data events inferred.
 Generation of sufficient 
background statistics to demonstrate adequate rejection power can be problematic, 
especially for highly ``signal-like'' background events, which are the 
hardest to reject.
\end{sloppypar}


Ultimately, 
separation of neutrino events from backgrounds is all about hypothesis testing -
given an observed event, 
which of a series of possible origins  (downgoing muons, 
upgoing neutrinos) is the most likely cause of the event? In the conventional
method, we  assess this by firstly reducing the data set through a maximum likelihood
reconstruction, then by  comparing the remaining
 data to Monte Carlo signal and background
expectations, using the Monte Carlo to predict how many events of a given type
are expected from both signal and background. At the reconstruction level we expect that
the event sample is still background dominated. After strong quality cuts, we expect the
sample to be signal dominated. In this paper, I describe how Bayesian inference
can be used to take account of prior knowledge of the expected signals and
backgrounds at the  reconstruction stage, leading to much greater initial background
rejection than the conventional method. Finally, I briefly indicate how
integration of posterior probability densities can
be used to directly assess the probability that an event is either signal or
background, providing a single cut parameter signal/background separation.
\section{Bayesian approach to event reconstruction}
In this section we first examine  the conventional maximum likelihood
 approach 
to the separation of signal events from background events in a neutrino
detector, and how the need to account for prior probabilities of
hypotheses leads naturally to the Bayesian method.
 In the maximum likelihood method, the likelihood function alone is 
used to assess the origin of an observed event, with the implicit
assumption that all hypotheses
are a priori equally probable. We then propose 
 a Bayesian  approach to event reconstruction, 
where the prior knowledge of the probabilities of the hypotheses is combined
with the likelihood function in assessing the most likely origin of
an event.

\subsection{The maximum likelihood method}
\label{3d}
A neutrino detector \cite{B4} gives us information
 about the detected muon event in the form of event
observables such as photon
arrival times and numbers of detected photons per optical module. The likelihood 
function $\mathcal{L}(E \mid H)$ describes the probability that any of the
events from the total possible event set $\left\{E\right\}$ come from any
of the possible set of hypotheses $\left\{H\right\}$. 
 These hypotheses could include for example
single muons, multiple
muon bundles, uncorrelated muons from separate cosmic ray primary events,
or upward going neutrino induced muons.
As an example, if we have
 a detector that measures photon arrival times with only a Gaussian timing
error due to the photomultiplier tubes
 (the case for a non-scattering medium), the likelihood minimisation
reduces to the familiar $\chi^2$ minimisation. 
Usually in a neutrino detector the assumption is made that the hypotheses to
be tested are single muons. The maximum likelihood method is used and the 
hypothesis space searched until the muon track hypothesis with the highest
likelihood is found. In what follows we will assume we have a minimisation 
algorithm that will always find the global minimum in the hypothesis space. 

The events are reconstructed, and assigned their best fit parameters, one of
which is the arrival direction, $\theta$. Since we are interested in upgoing 
muons from neutrinos, a cut is made to keep 
only events with $\theta > 90^{\circ}$. What we can infer about the origin
of these events
depends on what type of detector we have. There are three scenarios - 
\begin{enumerate}
\item {\bf The ideal detector}
Suppose the detector consisted of a billion optical modules on a one metre
lattice instrumenting a cubic kilometre of detector medium. The event information
would be so abundant and therefore the hypotheses constrained so
 strongly that there would be no doubt that the reconstructed direction
was in fact the true direction of the originating true hypothesis $H_t$.
Mathematically,
\begin{eqnarray}
    \mathcal{L}(E \mid H) &\not=& 0 \;\;  H = H_t \\
                          &=& 0 \;\;  H \not= H_t  
\end{eqnarray} that is, the likelihood would only be non-zero for the true 
hypothesis.

 Monte Carlo simulations of such
a detector would reveal that the chance of any downgoing event 
misreconstructing into the upward direction would be negligible. 
\item {\bf The non-ideal detector}
At the other extreme we have the detector where little if any constraint is
placed on possible hypotheses, due to factors such as a sparsely instrumented
detector volume, or a scattering medium where the Cerenkov light cone is 
smeared out completely.
All hypotheses would have essentially the same likelihood 
\begin{equation}
         \mathcal{L}(E \mid H_i) \approx  \mathcal{L}(E \mid H_j) \;\;
 \forall \;\; i,j
\end{equation} and a reconstruction reveals nothing about the origin of the 
event. 
 In this case, the best fit directions for the events 
would be essentially random, and all one could really say about a given 
event is that it was most likely to be due to a background event, since these
overwhelm the signals completely. Again Monte Carlo simulations would confirm
the inability of such a detector to separate signal and background. 
\item {\bf The realistic detector}
The final case is the realistic detector. In practise, we try not to construct
non-ideal detectors, but can never afford to build 
the ideal detector. We are left with an in-between situation, where after
the maximum likelihood reconstruction we still have background events
mis-reconstructed as upgoing events, but our Monte Carlo simulations tell us
that signal can be separated out of the backgrounds by the application of
quality cuts to the events to isolate the best upgoing events and reject the
falsely reconstructed background events.
\end{enumerate} 

\subsection{Inferring the origin of
reconstructed events in ``realistic'' 
detectors}
\begin{figure}[h]
\begin{center}

\setlength{\unitlength}{1.0cm}
\begin{picture}(6,6)\thicklines

  \put(0.5,0.5){\framebox(2,5)}
  \put(4.5,2){\vector(-3,1){5.5}}
   \put(4.5,4){\vector(-3,-1){5.5}}

  \put(2.5,3){\circle{0.15}}
  \put(2.0,3){\circle{0.15}}
  \put(0.5,3){\circle{0.15}}

  \put(2.0,3){\circle{0.15}}
 

  \put(2.5,2.5){\circle{0.15}}
  \put(1.5,2.5){\circle{0.15}}

  \put(2.0,3.5){\circle{0.15}}
  \put(1.0,3.5){\circle{0.15}}
  \put(1.0,2.5){\circle{0.15}}
 

\put(2.7,1.5){\it Best upgoing hypothesis: $H_u$}
\put(2.7,1){\it Likelihood: $\mathcal{L}(E \mid H_u)$}
\put(2.7,0.5){\it Prior probability: $P(H_u)$}

\put(2.7,5.3){\it Best downgoing hypothesis: $H_d$}
\put(2.7,4.8){\it Likelihood: $\mathcal{L}(E \mid H_d)$}
\put(2.7,4.3){\it Prior probability: $P(H_d)$}

\put(0.7,3.8){\it Event $E$, }

\end{picture}

\caption{\label{recofig} {Event reconstruction in a neutrino detector. ``Best''
hypotheses refer to those that maximise the likelihood function  $\mathcal{L}(E \mid H)$ alone, 
without consideration of the prior probabilities $P(H)$.}}
\end{center}
\end{figure}
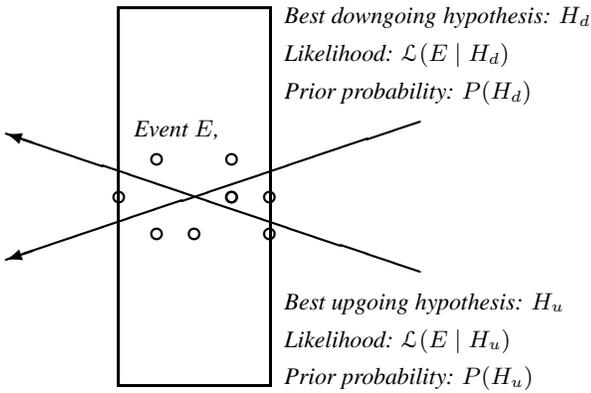

The conventional likelihood reconstruction method seeks to find the
hypothesis with the highest likelihood to have produced the observed
pattern of event observables. During the search through the hypothesis
space, all hypotheses are considered as equally probable, and therefore the
hypothesis that maximises the likelihood is taken as the best fit. For instance,
if the final choice is between an upgoing fit and a downgoing fit with only
a slightly smaller likelihood, the upgoing fit is chosen as the reconstructed
track. However, if one considers that the rate of expected downgoing
atmospheric muons is zenith dependent, and overall greater by many orders
of magnitude than the expected rate of upgoing neutrino-induced
muons, then it is not clear that the best choice  of the reconstructed direction
is upgoing. This situation is shown in figure \ref{recofig}. The event $E$ has
been reconstructed, and a best upgoing hypothesis $H_u$, with likelihood
 $\mathcal{L}(E \mid H_{u})$, has been found. The best downgoing hypothesis
is  $H_d$, with likelihood $\mathcal{L}(E \mid H_{d})$. The a priori
 known rates of
the hypotheses are denoted by the prior probabilities, $P(H_u)$ and $P(H_d)$. Depending on
the arrival direction of $H_d$, $P(H_d)$ can be orders of magnitude greater than $P(H_u)$.
These prior probabilities must
be taken into account in inferring the origin of the event. What we really
need to know is the {\em probability of the hypotheses, given the data}, denoted
 $P(H \mid E)$.  Bayes' theorem (Bayes, 1763) tells us that the {\em posterior} probability
 $P(H \mid E)$, is connected to the likelihood through the prior function 
\begin{equation}
\label{bayestheorem}
    P(H\mid E) = \frac{\mathcal{L}(E \mid H) P(H)}{P(E)}
\end{equation}
where 
\begin{equation}
     P(E) = \int \mathcal{L}(E \mid H) P(H) \mathrm{d}H
\end{equation}
Since $P(E)$ is a constant (an integral over all possible hypotheses), we
can more simply state that the most probable hypothesis is the one that
maximises the {\em joint} probability distribution $\mathcal{L}(E \mid H) P(H)$, or that
\begin{equation}
    P(H\mid E) \propto \mathcal{L}(E \mid H) P(H)
\end{equation}

Bayes' theorem lets us account for
the  prior probabilities of the hypotheses, whereas 
the conventional reconstruction, 
through an implicit assumption of the uniformity of the hypotheses, does not 
do this. 
Even if the likelihood of the upgoing hypothesis is the largest,
 if the prior probability of $H_d$ is sufficiently greater than that 
of $H_u$, such that 
\begin{equation}
 {\mathcal{L}}(E \mid H_{d}) P(H_d) >  {\mathcal{L}}(E \mid H_{u}) P(H_u)
\end{equation} 
then we must reject the event as downgoing. However, if after accounting for
the prior probabilities, the event appears to be upgoing, i.e. 
\begin{equation}
 {\mathcal{L}}(E \mid H_{u}) P(H_u) >  {\mathcal{L}}(E \mid H_{d}) P(H_d)
\end{equation} 
there might still be another
downgoing hypothesis  $H_a$ that satisfies
 \begin{equation}
 {\mathcal{L}}(E \mid H_{a}) P(H_a) >  {\mathcal{L}}(E \mid H_{u}) P(H_u)
\end{equation}
making $H_a$ the most likely origin of the event, forcing us to reject the event
as downgoing.  
 Cleary, in order to find the most likely origin of the event, we must find the hypothesis
$H$ that  maximises
the joint probability distribution \mbox{${\mathcal{L}}(E \mid H) P(H)$}.

We simply incorporate prior information into the reconstruction, by maximising
\mbox{${\mathcal{L}}(E \mid H)\;\;P(H)$}, instead of
\mbox{${\mathcal{L}}(E \mid H)$}.  A first attempt might be
 to make $P(H)$ a simple function of zenith angle, expressing the expected rates of 
downgoing atmospheric muons and neutrino-induced upgoing muons. During a 
search through the hypothesis space,
 downgoing hypotheses would be given more (zenith dependent) weight, 
 with the result that more events would reconstruct in the more
probable downward direction. This Bayesian reconstruction has been used in the
analysis of the AMANDA B10 data from the austral winter 1997 (Andres et al. 2001, DeYoung 2001,
 Wiebusch et al. 2001), where it has significantly improved the reconstruction level background
muon rejection, 
 resulting in a simpler analysis, requiring fewer subsequent quality cuts to separate the neutrino
signal.

Having seen the importance of the prior probability $P(H)$ in analysing the
data in a \emph{realistic} detector, we can now go back and see how
 Bayesian inference also explains the behaviour of the \emph{ideal} and 
 \emph{non-ideal} detectors discussed in section \ref{3d}, as the two extreme cases
of the influence of the prior probability.
 In the case of the \emph{ideal detector}
 the 
hypotheses are so strongly constrained that the likelihood is only non-zero for
 the true hypothesis $H_t$ and zero for all other hypotheses
\begin{eqnarray}
    \mathcal{L}(E \mid H) &\not=& 0 \;\;  H = H_t \\
                          &=& 0 \;\;  H \not= H_t  
\end{eqnarray} 
As long as $P(H_t)\not= 0$, the posterior probability $P(H\mid E)$ is
 only non-zero for the true hypothesis, and the detector constrains the hypotheses
 so strongly that prior knowledge of the  hypotheses is irrelevant. The
complete opposite is the case for the \emph{non-ideal detector}. The detector
places no constraint on the hypotheses, the likelihoods of all $E$ given all
hypotheses are essentially equal
\begin{equation}
         \mathcal{L}(E \mid H_i) \approx  \mathcal{L}(E \mid H_j) \;\;
 \forall \;\; i,j
\end{equation} 
and therefore 
\begin{equation}
      P(H\mid E) \propto P(H)
\end{equation}
Since the detector cannot constrain the hypotheses at all the only inference 
that can be made is based on the prior probabilities, ie each event is more
likely to be a background than a signal simply because backgrounds overwhelm
signals. 

As expected the \emph{realistic detector} lies in between the two cases
 discussed so far. Prior probabilities don't matter when analysing the ideal 
detector, but in the non-ideal detector prior probabilities tell us all we can know
about an event. For the realistic detector, both the likelihood and the 
prior probability must be used when inferring the origin of an event.


Using this Bayesian  reconstruction of course still only finds the hypothesis
with the maximum value of \mbox{${\mathcal{L}}(E \mid H) \;\; P(H)$}.
The probability that a class of hypotheses (for instance any upgoing hypothesis)
 was the cause of the event is found by integrating the posterior probability
for that class. In the next section we discuss how integrating these posterior
probabilities separately over signal and background hypotheses allows us to 
directly calculate the probability that any event is a signal or a background.

\section{Bayesian separation of signal and backgrounds using integrated 
posterior probabilities}
In this section we briefly discuss the Bayesian method for assessing the probability
that an event is a signal or a background. Elsewhere (Hill, 2001) these ideas are 
developed in greater detail. 
As our starting point, we restate Bayes' theorem (equation \ref{bayestheorem}) for the
posterior probability density for hypothesis $H$ as the origin of event $E$:

\begin{equation}
    P(H\mid E) = \frac{\mathcal{L}(E \mid H) P(H)}{P(E)}
\end{equation}
where 
\begin{equation}
     P(E) = \int \mathcal{L}(E \mid H) P(H) \mathrm{d}H
\end{equation}
is the event probability density function (p.d.f.).
We split the hypothesis space up into separate classes of signal and
background hypotheses. For example, the background hypotheses could be single minimum 
ionising muons, muons undergoing catastrophic energy losses (e.g. brems\-strahlung), 
mutiple muon
bundles, or overlapping events. The signal hypotheses could be atmospheric
neutrino induced muons, which travel upward through the earth. 

The prior probabilities of these hypotheses
are denoted $P(S)$ and $P(B)$.
  The functions ${\mathcal{L}}(E \mid S)$ and
${\mathcal{L}}(E \mid B)$ describe the likelihood of a given hypothesis to produce a set of detector
observables (eg photon arrival times,
 photoelectron densities, time-over-thresholds). For completeness, the set
of events $\left\{E\right\}$ includes every type of event we could observe in the
detector, right down to events where only detector noise contributes, and
there is no contribution from, for example, a muon.

Then Bayes' theorem for the posterior p.d.f. of the signal hypotheses is
\begin{equation}
   P(S\mid E) =  \frac{\mathcal{L}(E\mid S) P(S)}{P(E)}
\end{equation}
and for the background hypotheses 
\begin{equation}
   P(B\mid E) =  \frac{\mathcal{L}(E\mid B) P(B)}{P(E)}
\end{equation}
where the event p.d.f. is
\begin{equation}
  P(E) = \int \mathcal{L}(E\mid S) P(S) \mathrm{d}S +
            \int \mathcal{L}(E\mid B) P(B) \mathrm{d}B
\end{equation}
The probabilty $P_s$ that \emph{any} of the signal hypotheses was the cause of $E$
is found by integrating over the posterior p.d.f. $P(S\mid E)$ to give
\begin{equation}
   P_s = \int  P(S\mid E) \mathrm{d}S = \frac{\int \mathcal{L}(E\mid S) P(S)
                                    \mathrm{d}S}{P(E)}
\end{equation}
and likewise for backgrounds
\begin{equation}
   P_b = \int  P(B\mid E) \mathrm{d}B = \frac{\int \mathcal{L}(E\mid B) P(B)
                                    \mathrm{d}B}{P(E)}
\end{equation}
 The ratio
\begin{equation}
   \frac{P_s}{P_b} =  \frac{\int P(S\mid E)  \mathrm{d}S }{\int P(B\mid E) \mathrm{d}B} = 
             \frac{\int\mathcal{L}(E\mid S) P(S)
        \mathrm{d}S}{\int\mathcal{L}(E\mid B) P(B) \mathrm{d}B}  
\end{equation}
gives the ``betting odds'' in favour of a signal origin over a background
origin for any event.

 Also since
\begin{eqnarray}
     P_s + P_b = 1
\end{eqnarray}
either $P_s$ or $P_b$ gives a complete description of the probability of
an event to be either signal or background and therefore serves as a single
cut parameter.

We could, in principle, calculate the event probability for every observed
event. In practice, this would be too time prohibitive to be of practical use.
A better solution would be to initially use the Bayesian reconstruction to
greatly reduce the data set, then calculate the integrated posterior signal
and background 
probabilities for this small subset of events. 

\section{Conclusions} 
In this report we have addressed, from a Bayesian perspective,
 the two critical  questions faced when analysing neutrino telescope
data - what is the most likely arrival direction of an event, and what
is the probabilty that subsequent upwardly reconstructed 
events are really upgoing signal and not 
mis-reconstructed downgoing atmospheric muons? To answer both questions, we
have applied Bayes' theorem, which assesses the plausibility of hypotheses
given observed data
by combining prior knowledge of the hypotheses with the likelihood
function. In the reconstruction
case this results in more events reconstructing in the more probable 
downward direction. In the case of
background rejection, we calculate the probability that a given
event was upgoing by integrating over posterior probabilities for all 
possible upgoing and downgoing hypotheses.

\begin{acknowledgements}
I thank all my colleagues in the AMANDA experiment 
for many lively discussion sessions on these topics. I wish
to particularly thank T. DeYoung, S. Hundertmark, J. Kim, D. Steele,
 A. Karle, K. Rawlins, R. Morse and F. Halzen
 for their discussions and suggestions. 
\end{acknowledgements}

\clearpage

\end{document}